\documentclass[conference]{IEEEtran} 
\usepackage{amsmath}
\usepackage[switch]{lineno}

\usepackage{graphicx}
\usepackage{subfigure}
\usepackage{multirow}
\usepackage{cite}
\usepackage{color}

\usepackage{tabularx}
\usepackage{balance}
\usepackage{soul}

\usepackage{amssymb}

\title{Neuromorphic adaptive edge-preserving denoising filter}

\author{\IEEEauthorblockN{Aidana Irmanova, Olga Krestinskaya, and Alex Pappachen James }
\IEEEauthorblockA{Department of Electrical and Electronics Engineering\\
Nazarbayev University,
Astana\\
Email: http://www.biomicrosystems.info/alex}}

\begin{document}

\maketitle

\begin{abstract}
In this paper, we present on-sensor neuromorphic vision hardware implementation of denoising spatial filter. The mean or median spatial filters with fixed window shape are known for its denoising ability, however, have the drawback of blurring the object edges. The effect of blurring increases with an increase in window size. To preserve the edge information, we propose an adaptive spatial filter that uses neuron's ability to detect similar pixels and calculates the mean. The analog input differences of neighborhood pixels are converted to the chain of pulses with voltage controlled oscillator and applied as neuron input. When the input pulses charge the neuron to equal or greater level than its threshold, the neuron will fire, and pixels are identified as similar. The sequence of the neuron's responses for pixels is stored in the serial-in-parallel-out shift register. The outputs of shift registers are used as input to the selector switches of an averaging circuit making this an adaptive mean operation resulting in an edge preserving mean filter. System level simulation of the hardware is conducted using 150 images from Caltech database with added Gaussian noise to test the robustness of edge-preserving and denoising ability of the proposed filter. Threshold values of the hardware neuron were adjusted so that the proposed edge-preserving spatial filter achieves optimal performance in terms of PSNR and MSE, and these results outperforms that of the conventional mean and median filters.

\end{abstract}

\section{Introduction}

\IEEEPARstart{T}{he} neurons and neural networks connected with the visual pathways of visual cortex exhibit several important properties for intelligent image processing under highly noisy environments \cite{v1,v2}. The ability for the neurons to respond to input stimuli, its ability to ignore noise by learning from repeated stimuli over a period of time is primary to its adaptive behavior. Once the neuron learns an input stimuli, its response to a known and unknown inputs stimuli can be differentiated by the strength of its output responses. This behavior can be modeled as weighted addition of inputs over a period of time followed by threshold operations serving as activation function for neural firing \cite{rojas2013neural}. Since the neuron fires based on only the known or learned stimuli, they serve as a natural system of similarity detection between two stimuli - comparing an input stimuli with that of the learned stimuli in real-time \cite{v3}. In this paper, we exploit this ability of the neuron to determine the similarity between two pixels for edge preserving image filtering operations.

There is a class of image filtering techniques that rely on similarity calculations between the pixels to extract and preserve structural information. One such effective technique is G-neighbour filtering \cite{gn}, that uses the the similarity between the neighborhood pixels within the filtering window to reject the dissimilar pixels from the computation of the filtering operations. In this study, we base our focus on mean filtering operation as it results in lose of edge information with an increase in filter window size.

Image denoising has been a well-studied question in the image processing field and continues to attract researchers with an aim to perform better restoration. As the the number of pixels per unit area of a chip is continuously increasing, modern image capturing devices are increasingly sensitive to noise \cite{chatterjee2010denoising}. Therefore, built-in image denoising filters are required to reduce the present noise in resultant image.

There are several methods proposed for hardware implementations of such image denoising filters \cite{fpga}, \cite{sekanina2002image}. Conducted work shows that image denoising on hardware level provide fast execution and is well suited for real time image processing. But to build such hardware filters, denoising algorithms needs to be less complex and non-iterative\cite{vinh2008fpga}

In this paper, we propose using neurons for detecting the similarity between the central pixel to its neighborhood pixels within the filtering window. Similarity scores detected by neuron are used as mask matrix that builds up an adaptive denoising filter. Binary weights of the mask or similarity scores are set by the threshold value of the neuron, which equals to maximum possible difference between pixels. To measure effectiveness of proposed denoising algorithm quantitative performance measures such as peak signal-to-noise ratio (PSNR) and mean square error (MSE) are calculated. Visual quality evaluation of the images are also conducted in system level simulations. In this paper Gaussian Noise was added to test the performance of proposed denoising method.

\begin{figure*}[!ht]
\centering

\includegraphics[width=150mm]{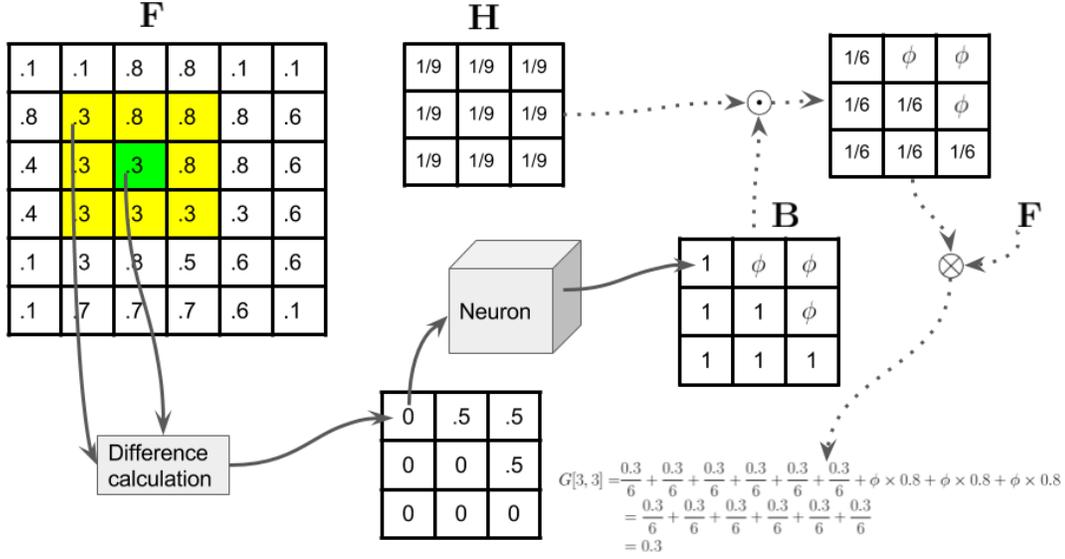}
\caption{An example illustration of the modification of the spatial filtering operation by introduction of a binary similarity mask $\mathbf{B}$ that modifies the filter mask $H$. The elements of $\mathbf{B}$ are determined using a neuron circuit proposed in this paper. }
\label{example}
\end{figure*}

\section{Proposed Filter}

Traditional image filtering operations $F[x,y]$ and its filter mask $H[x,y]$ can be represented as:

\begin{align}
 G[x,y] &= \mathbf{F}[x,y] \otimes  \mathbf{H}[x,y]\\ 
 &= \sum_{i= -w/2}^{w/2} \sum _{j= -h/2}^{h/2} \mathbf{F}[x+i,y+j]\mathbf{H}[i,j]
\end{align}
where, $w$ and $h$ is the width and height of the image window or the mask. We modify this filtering operation by including a binary similarity mask $\mathbf{B}$ that has either null value $\phi$ or 1 and changing the values of $\mathbf{H}$, for example changes to mean filter mask with values $1/n$, where $n$ is the number of non-null values in the mask. This modification is formally introduced as:

\begin{align}
 G[x,y] &= \mathbf{F}[x,y] \otimes  (\mathbf{H}[x,y]\odot \mathbf{B}[x,y])\\ 
 &= \sum_{i= -w/2}^{w/2} \sum _{j= -h/2}^{h/2} \mathbf{F}[x+i,y+j]\mathbf{H}[i,j]\mathbf{B}[i,j]
\end{align}
where, a multiplication of $\phi$ with any real number $\mathbb{R}$ results in null $\phi$, and those pixels are excluded from the filtering operation. The filter falls under the subclass of G-neighbor filter, where $\mathbf{B}$ serves as the similarity matrix representing the similarity between the center pixel to that of its neighborhood.










\begin{figure*}[!ht]
%
\centering
\includegraphics[width=150mm]{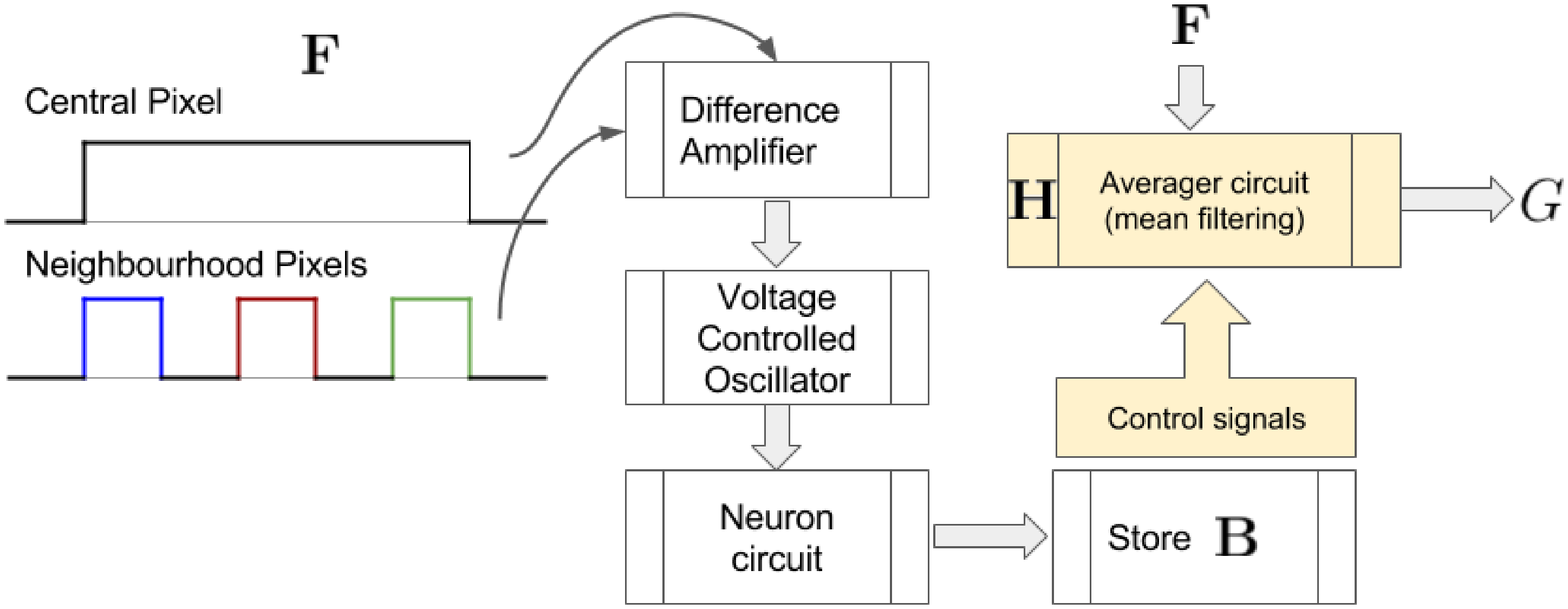}
\caption{The block diagram shows the various functional circuit block of the proposed system for a mean filter implementation.}
\label{blockdia}
\end{figure*}

Figure \ref{example} shows a numerical example demonstrating the functionality and working principle of the proposed spatial filtering operator. In this example, the mean filter mask is modified by using the similarity mask $\mathbf{B}$ and removes the dissimilar pixels from the filter output $G$ calculations.

\section{Hardware implementation}




The block diagram of proposed hardware implementation of adaptive mean filter is shown in Fig. \ref{blockdia}. The pixel in the center of the window is compared to its neighbor pixels sequentially in time using differential amplifier. Obtained analog differences are used as control signal to the voltage controlled oscillator (VCO) that compares this control voltage to an inbuilt sinusoidal signal. VCO produces signal spikes of different frequencies depending on the amplitude of the applied inputs. After VCO conversion, produced set of spikes is applied as input signal to the neuron that fires only for a tuned similarity threshold. Binary output $\mathbf{B}$ of the neuron, represented as logic high "1" in case of similar pixels and logic low "0" otherwise, is saved to Series-In-Parallel-Out (SIPO) shift register. Values saved in SIPO builds up the mask that controls the switches in the averaging circuit, activating only similar pixels for implementing mean operation.  

\begin{figure}[!ht]
\centering

\includegraphics[width=7cm]{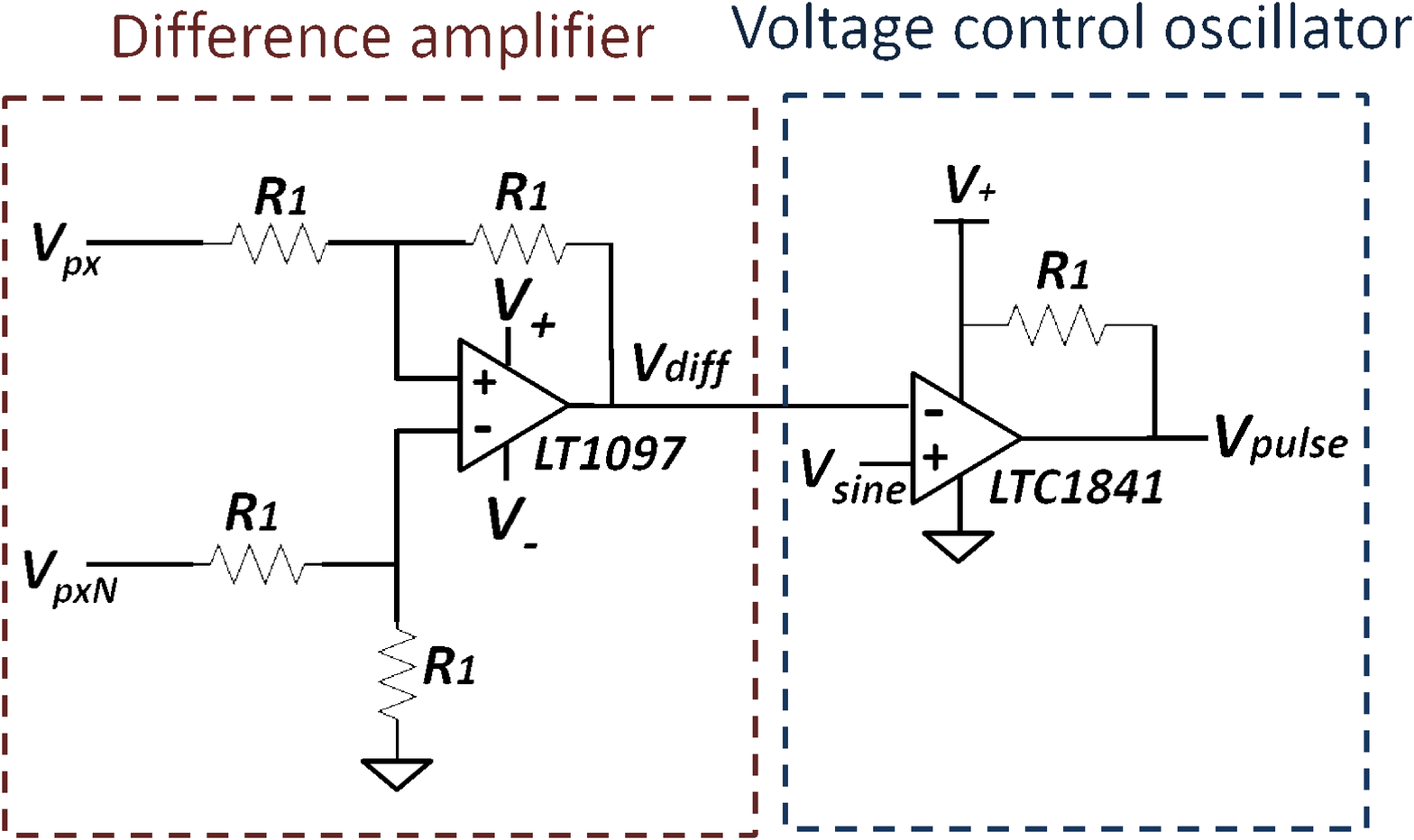}
\caption{Pre-neuron stage}
\label{1}
\end{figure}

The process of filtering noise within a single window can be divided to three stages: pre-neuron stage, that converts the pixel differences to the chain of pulses, neuron processing stage, that outputs mask values, and mean calculation stage, that provides filter output $V_{diff}$.
The analog circuit for the pre-neuron stage, consisting of the difference amplifier and VCO, is shown in Fig.\ref{1}. The difference amplifier calculates the difference of neighbor pixels and is based on LT1097 amplifier model. VCO, on the other hand, creates a chain of pulses of a particular frequency depending on the input voltage supplied from the amplifier. VCO is based on LTC1841 operational amplifier model and has two input signals to the comparator: the differential amplifier output and sine wave signal with an amplitude of $V_{sine}$. The increase of the $V_{diff}$ signal leads to the decrease of the duty cycle of the output pulses from the VCO.

\begin{figure*}[!ht]
\centering

\includegraphics[width=16cm, height=6cm]{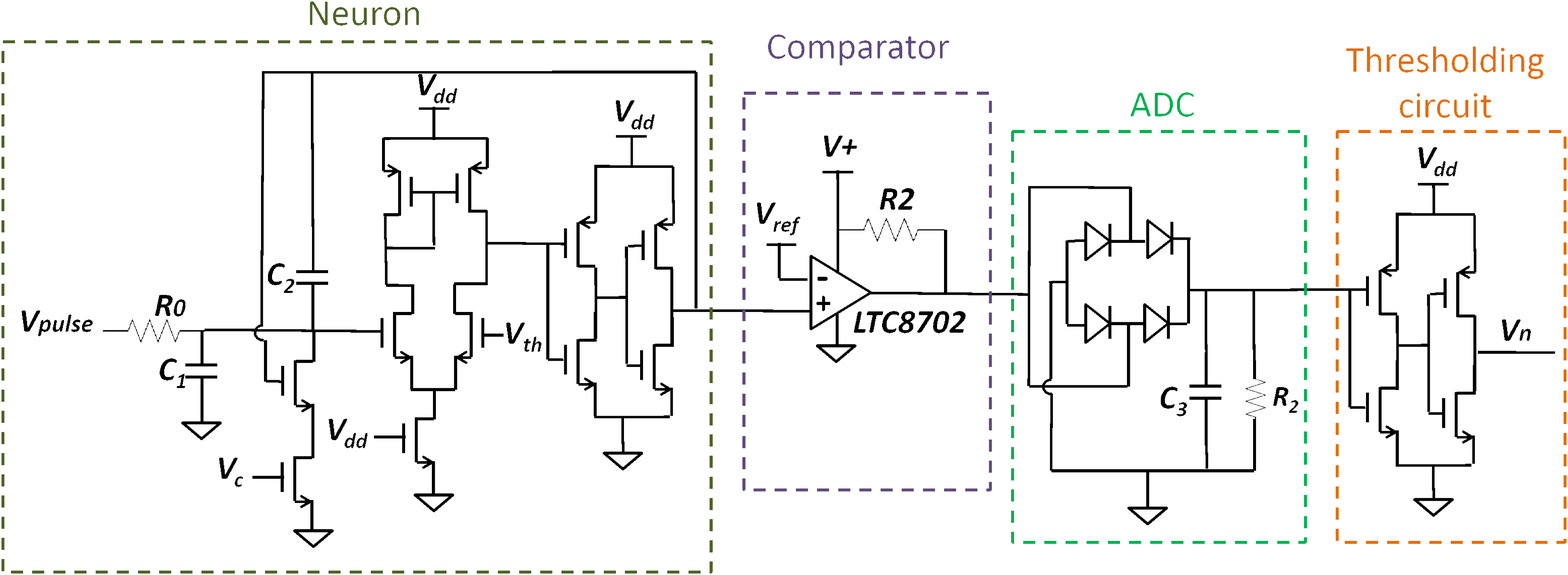}
\caption{Neuron design with four different stages for generating the desired pulse response.}
\label{2}
\end{figure*}

As regards neuron design, it was inspired from the model presented in \cite{sun2015cmos}, characterized by steep depolarization and repolarization phases. Figure \ref{2} shows modified circuit of the neuron, that consists of 4 blocks representing different functional characteristics of the neuron. The first block is the polarization or charging functionality of the neuron, which is realized through different levels of $C_1$ and $C_2$ capacitor values and CMOS transistors. The second block consists of single comparator  LTC8702 and acts as an activation function, which output will be chain of spikes, in case the neuron is charged to the $V_{ref}$ level. The amplitude of $V_{pulse}$ and the values of the capacitances impact the time required to make the neuron fire. If the difference of compared pixels are less than the firing threshold, the pre-firing time increases.

\begin{figure}[!ht]
\centering

\includegraphics[width=9cm]{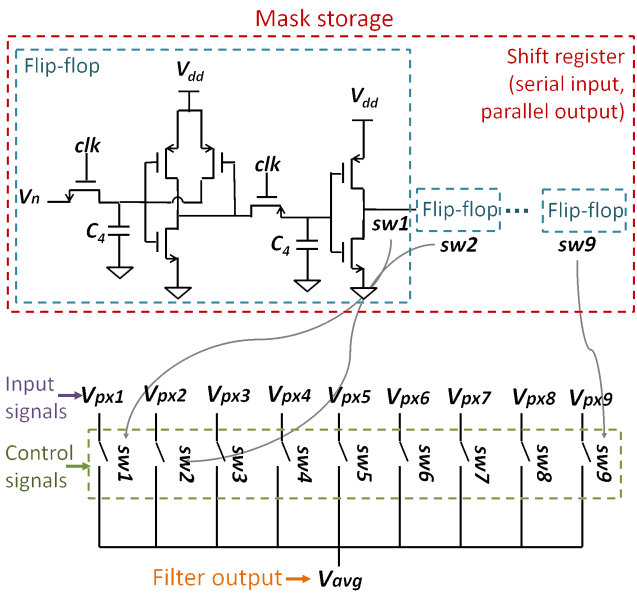}
\caption{Storing, control circuit and filter output}
\label{3}
\end{figure}

The following two neuron blocks are responsible for normalization the output spikes for further storage in the SIPO shift register. First, chain of pulses that neuron produces in case of firing are converted to DC voltage. Next, this signal is normalized to $V_{dd}$ amplitude and fetched to SIPO shift register shown in Fig. \ref{3}. SIPO shift register performs the storage functionality to preserve the complete mask for implementing mean operation and consists of 9 flip-flops. The neuron output $V_{n}$ is fetched to the first flip-flop. The output of the flip-flop is plays the role of input signal to the subsequent flip-flop to ensure the shift register functionality. The outputs of all flip-flips are read at the same time when the last flip-flip is activated. The output of each flip-flop is then fetched to control the corresponding switch in the averaging circuit in Fig. \ref{3} and the mean of activated pixel branches are calculated. 
The simulation of the proposed filter design was conducted in Spice. Configuration of the proposed filter circuit is provided in Table \ref{circ}.



\begin{table}[!ht]
\centering
\caption{Circuit configuration}
\label{circ}
\begin{tabular}{|l|l|}
\hline
NMOS, W/L ($\mu$m)                			& 0.36/0.18    \\ \hline
PMOS, W/L ($\mu$m)                			& 0.72/0.18    \\ \hline
$V+/V-$                    		  			& 3V/-3V         \\ \hline
$V_dd$                            			& 1V           \\ \hline
$V_{ref}$                         			& 1.12V        \\ \hline
$V_c/V_{th}$                     		 	& 0.6/0.4      \\ \hline
$V_{sine}$ (amplitude, frequency) 			& 3V, 100kHz            \\ \hline
clk (amplitude, frequency, duty cycle)      & 1V, 50Hz,2.5\%            \\ \hline
C1/C2 (pF)                      			& 1200/12000 \\ \hline
C3 ($\mu$F) 								& 1  \\ \hline
C4 (pF)										& 0.01        \\ \hline
$R_0/R_1/R_2$ ($\Omega$)                    & 1k/10k/100k  \\ \hline
\end{tabular}
\end{table}


\section{Results and Discussion}

\subsection{System level simulation}
\begin{figure}[!ht]
    \centering

      \subfigure[]{
    \includegraphics[height=3cm]{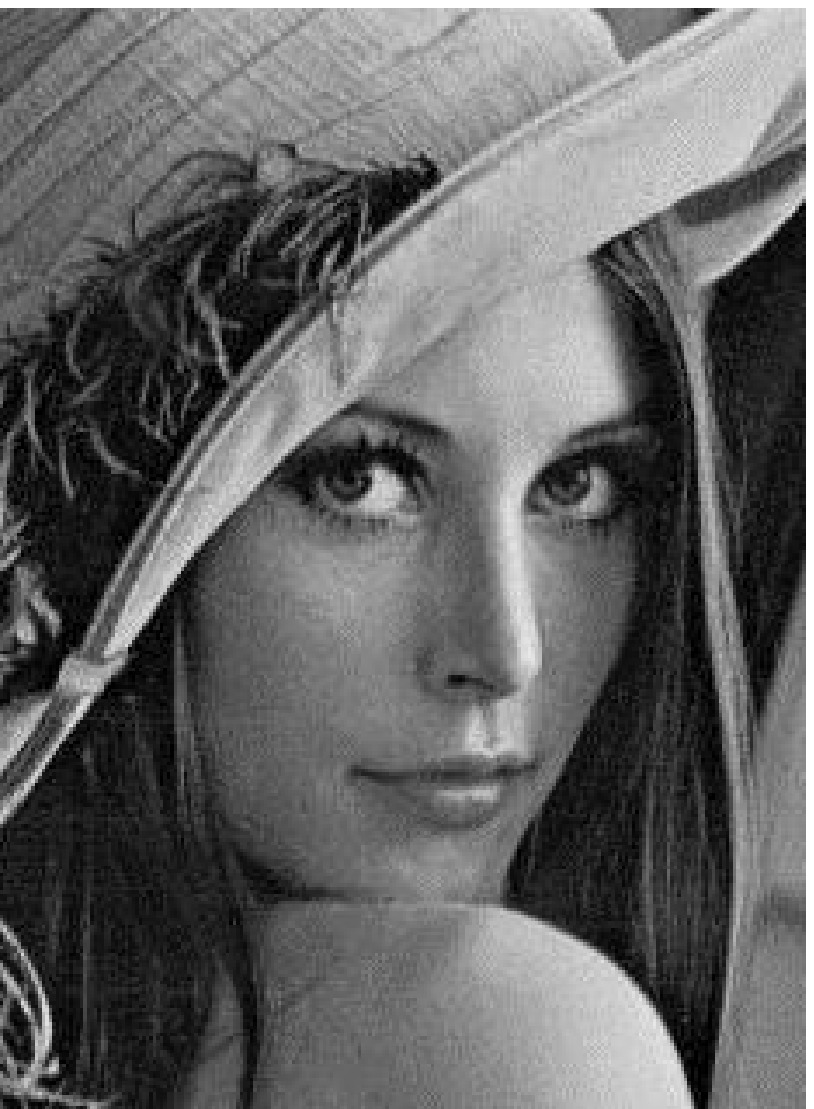}}
      \subfigure[]{
    \includegraphics[height=3cm]{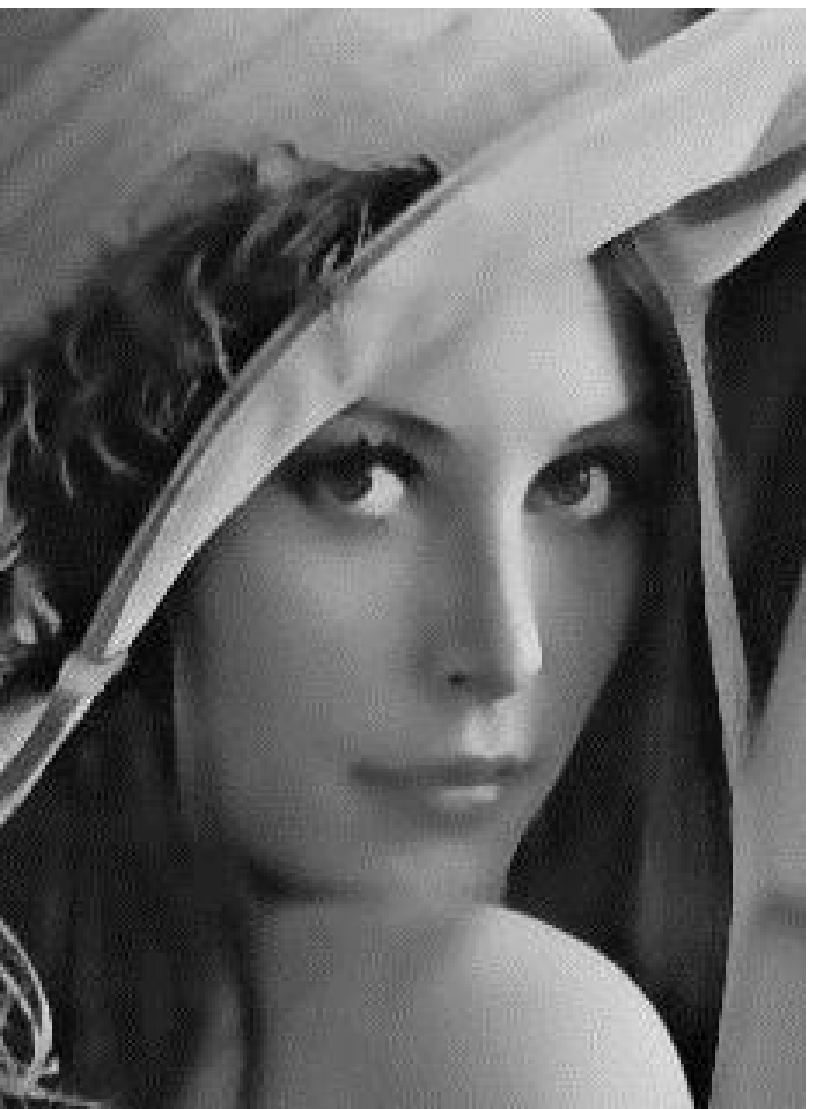}}
     \subfigure[]{
 
    \includegraphics[height=3cm]{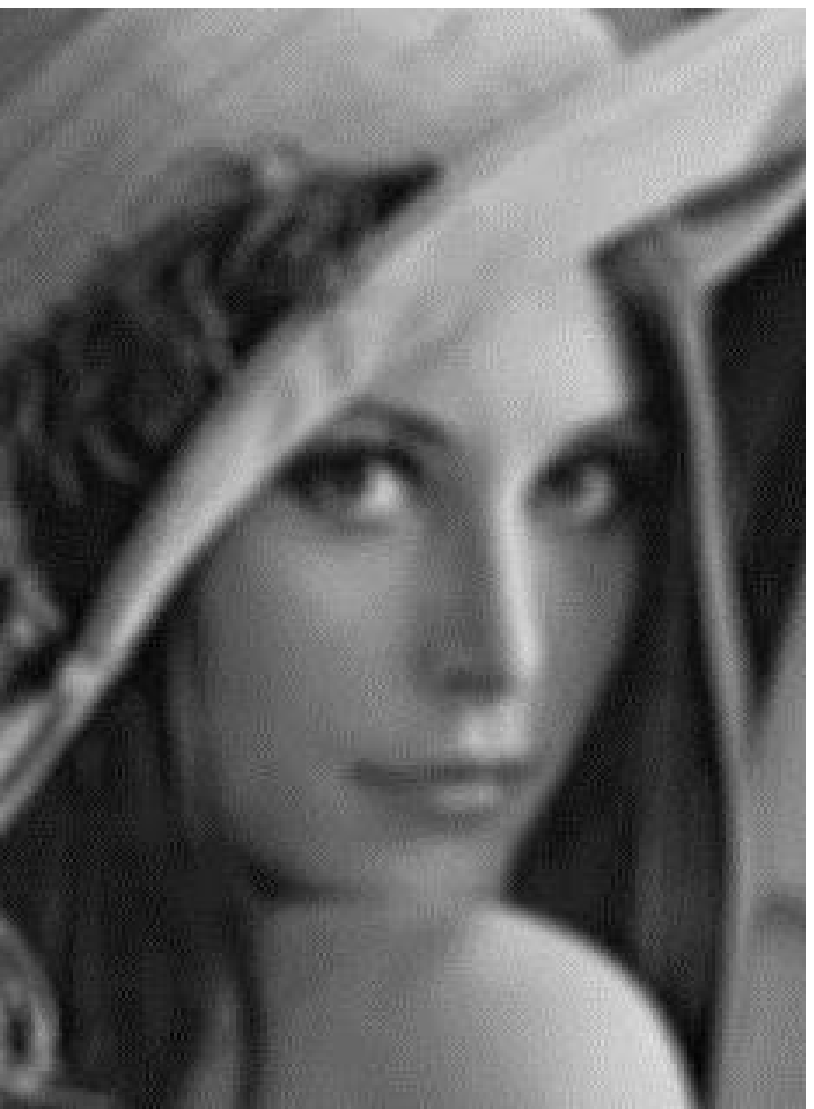}}
 
    \caption{Comparison of image filtering results:(a) initial picture, (b) application of adaptive mean and (c) conventional mean filter}
       \label{imres1}
\end{figure}

System level simulation of the proposed filter was  verified in MATLAB using 150 images from Caltech   database\cite{caltech101_2017} and adding Gaussian Noise at different rates(0.02 and 0.04). To achieve the optimum performance of the filter, simulation results of denoising the images with proposed method at different threshold levels were compared with the results of the conventional Mean filter. The quantitative performance of the filters was verified by peak signal to noise ratio (PSNR) and mean square error (MSE). Table \ref{my-table} presents the average PSNR and MSE values for all the simulated images for different threshold values. The optimum performance of the filter is achieved at the threshold of $\theta=0.3$ for both 0.02 and 0.04 rate of Gaussian noise. Comparing to the conventional filtering, MSE is reduced by 36.6\% and PSNR is increased by 2.5\%. This optimum threshold value is used in the hardware implementation of the proposed neuron-based adaptive G-neighbor filter. The comparison of the conventional mean filter and the proposed adaptive mean filter performances is presented in Fig. \ref{imres1}. The proposed neuron-based G-neighbor filter implementation outperforms the conventional mean filtering in terms of the noise reduction and preservation the image quality.

\begin{table}[]
\centering
\caption{Simulation results of adaptive and conventional mean filtering}
\label{my-table}
\begin{tabular}{|c|c|c|c|c|c|}
\hline
\multirow{2}{*}{\textbf{Noise rate}} & \multirow{2}{*}{\textbf{Threshold}} & \multicolumn{2}{c|}{\textbf{PSNR}}                                                         & \multicolumn{2}{c|}{\textbf{MSE}}                                                          \\ \cline{3-6} 
                                     &                                     & \textbf{\begin{tabular}[c]{@{}c@{}}Adapative\\ Mean\end{tabular}} & \textbf{Mean}          & \textbf{\begin{tabular}[c]{@{}c@{}}Adaptive\\ Mean\end{tabular}} & \textbf{Mean}           \\ \hline
\multirow{3}{*}{0.02}                & 0.2                                 & 36.75                                                             & \multirow{3}{*}{36.17} & 0.0029                                                           & \multirow{3}{*}{0.0043} \\ \cline{2-3} \cline{5-5}
                                     & 0.3                                 & 37.47                                                             &                        & 0.0022                                                           &                         \\ \cline{2-3} \cline{5-5}
                                     & 0.4                                 & 37.25                                                             &                        & 0.0025                                                           &                         \\ \hline
\multirow{3}{*}{0.04}                & 0.2                                 & 36.05                                                             & \multirow{3}{*}{35.63} & 0.0040                                                           & \multirow{3}{*}{0.0052} \\ \cline{2-3} \cline{5-5}
                                     & 0.3                                 & 36.54                                                             &                        & 0.0033                                                           &                         \\ \cline{2-3} \cline{5-5}
                                     & 0.4                                 & 36.39                                                             &                        & 0.0035                                                           &                         \\ \hline
\end{tabular}
\end{table}

\subsection{Hardware level simulation}

\begin{figure}[!ht]
    \centering

      \subfigure[]{
    \includegraphics[width=7cm]{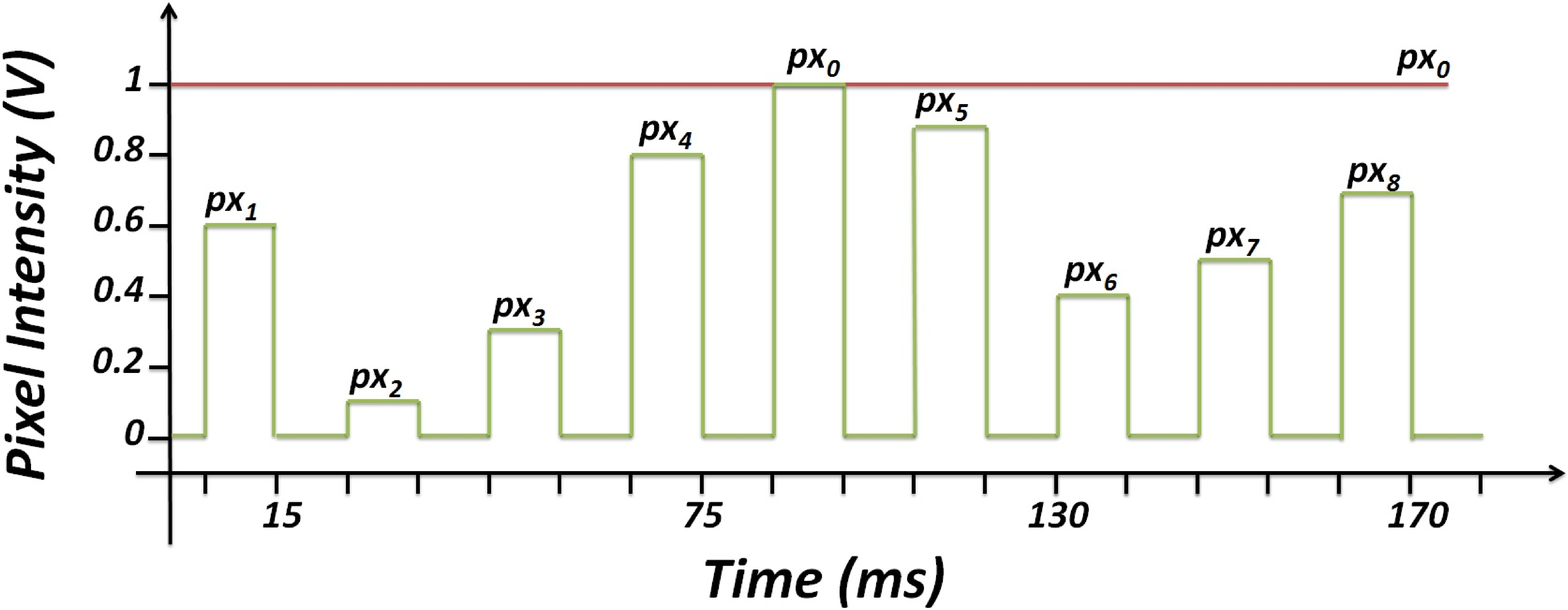}}
      \subfigure[]{
    \includegraphics[width=7cm]{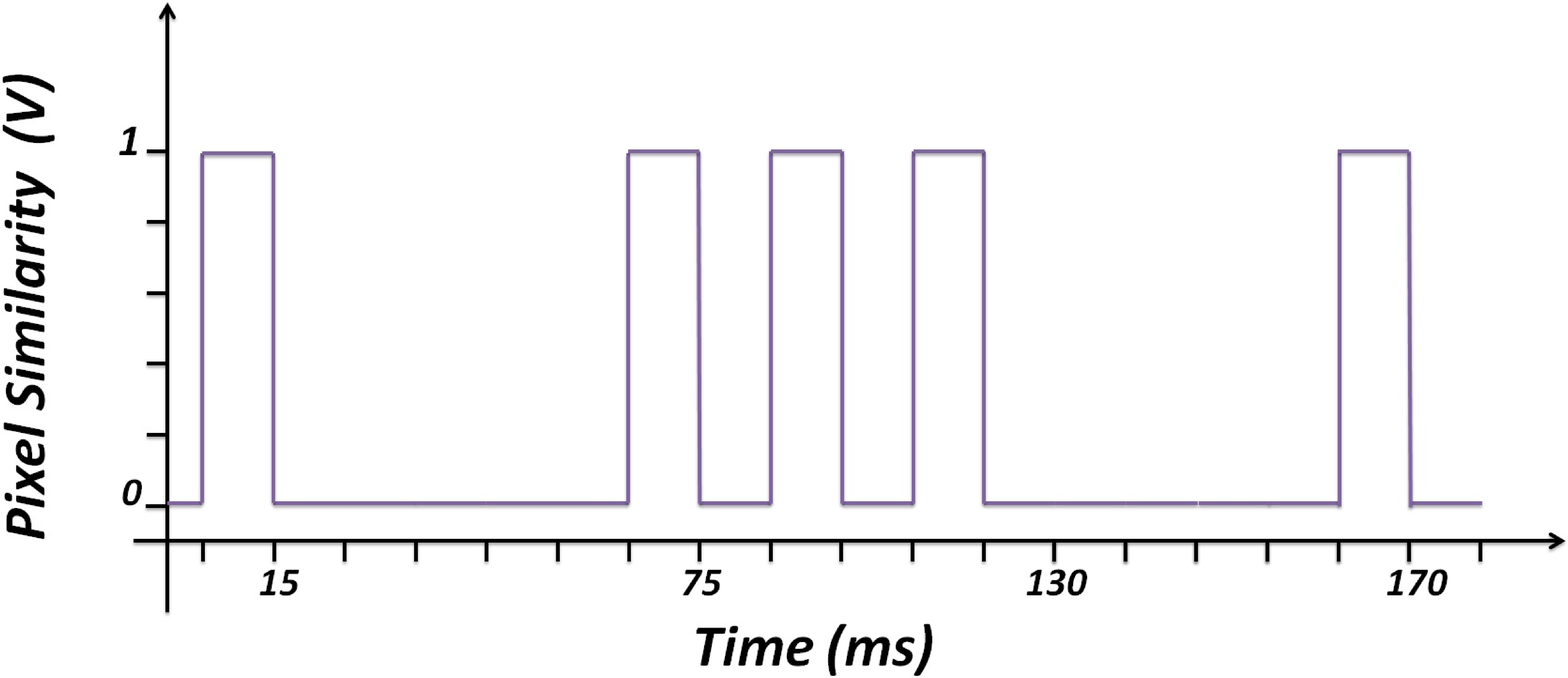}}
    \subfigure[]{
    \includegraphics[width=7cm]{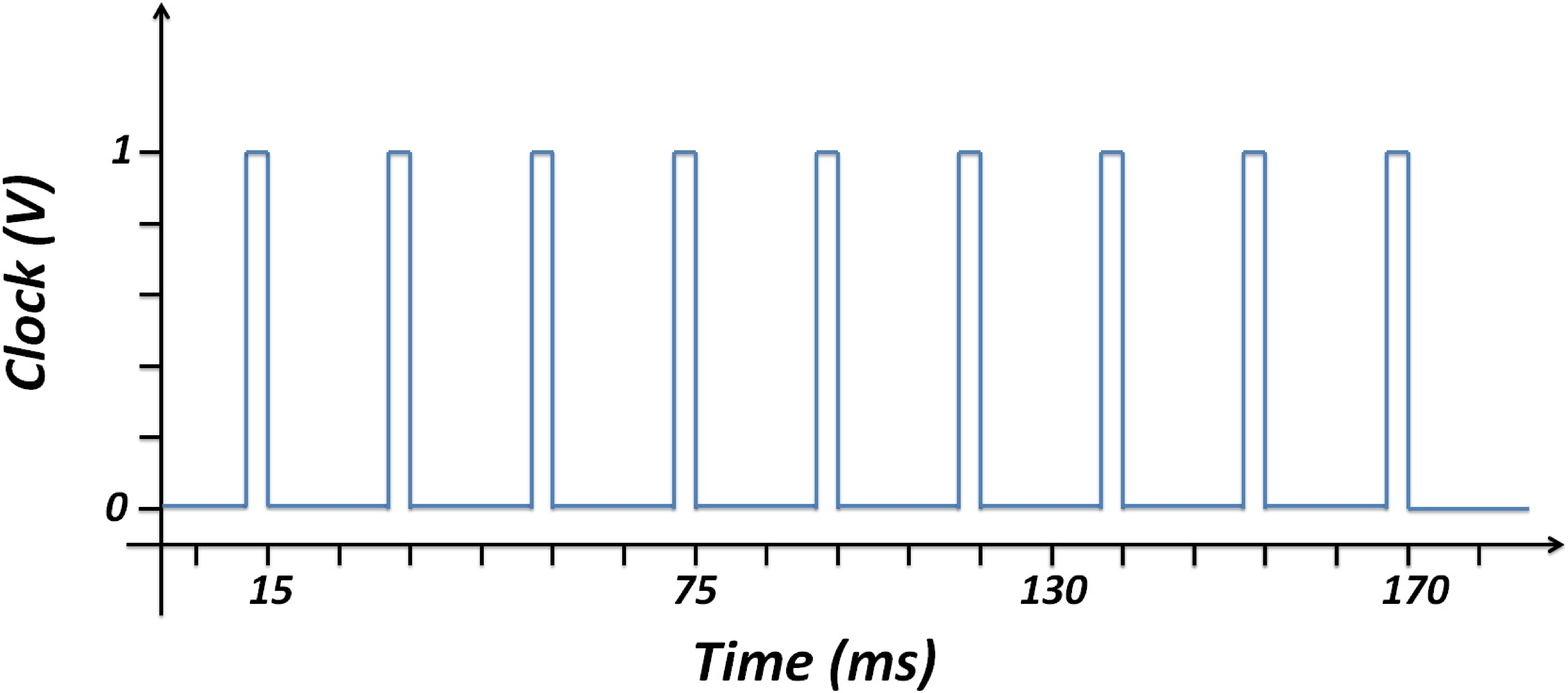}}
   \subfigure[]{
    \includegraphics[width=7cm]{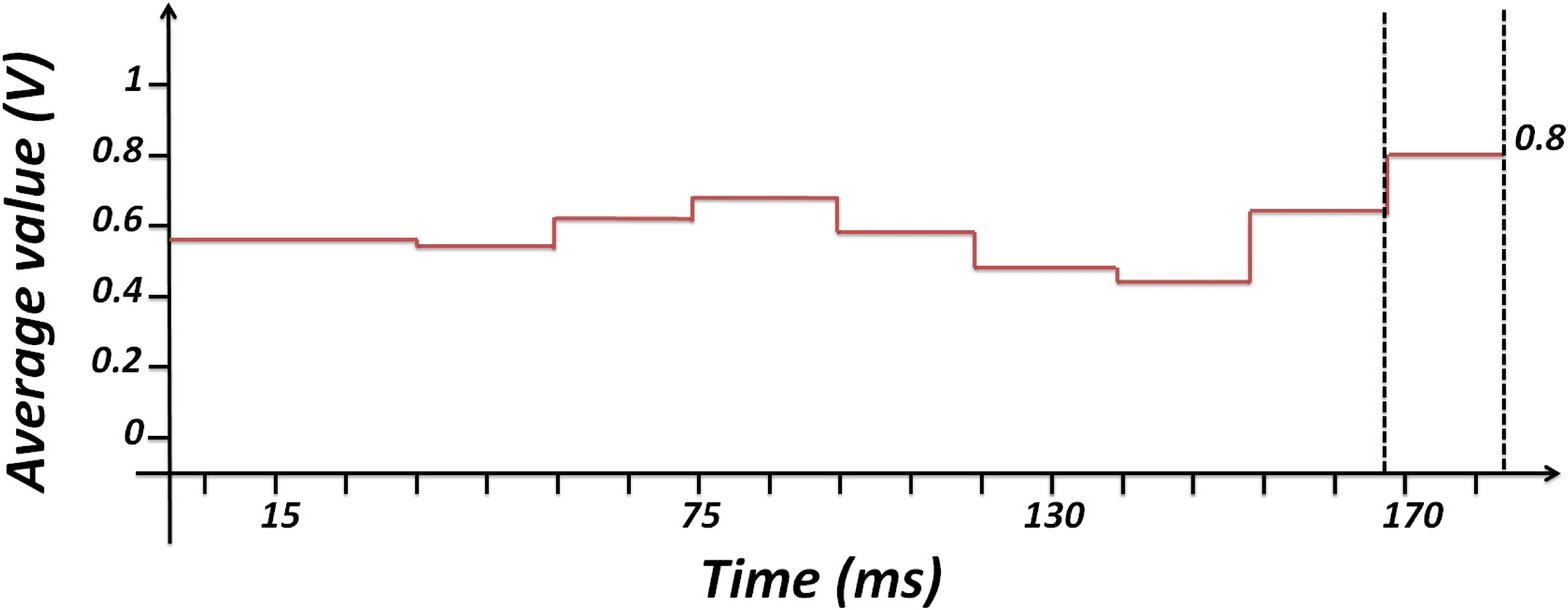}}

    \caption{Time diagram of (a) differential amplifier inputs (b) and corresponding neuron output signals, (c) clock signal and (d) output of averaging circuit}
       \label{td11}
\end{figure}

\begin{figure}[!ht]
\centering

\includegraphics[width=7cm]{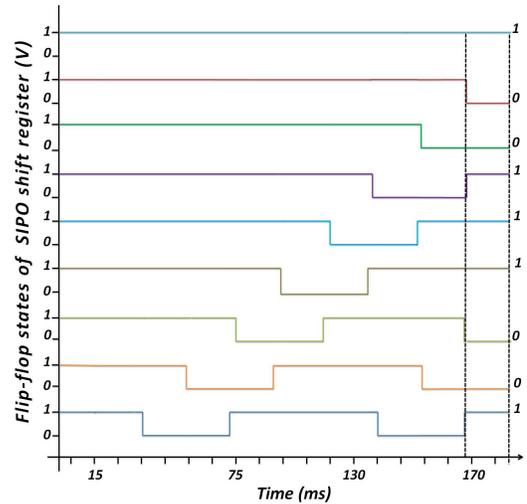}
\caption{Time diagram of Flip flop states of shift register}
\label{le}
\end{figure}
Proposed hardware implementation of adaptive mean filter exploits polarization property of the neuron reflecting its learning ability. However, the threshold level of neuron activation was set manually using results of system level simulations, which gives the ability of differentiating similar pixels. Nevertheless, design of the filter mimics the model of human visual cortex neuron, reacting to the chain of pulses fed as an input information and giving the output in binary values for building up the mask matrix that denoises the input signal preserving the information. An example for this is the human eye that can adapt to the noisy environment still detecting the edges and recognizing the objects in the image.

The simulation results of the hardware implementation of the proposed filter are shown in Fig.\ref{td11} and Fig.\ref{le}. Fig.\ref{td11} (a) illustrates the exemplar inputs to the filter. Signal denoted as $px_0$ refers to the value of the central pixel, while second input signal of $px_N$ amplitude corresponding N-th pixel within the window. Series output of the neuron for given example input signals is shown in Fig.\ref{td11} (b) while the activation of flip flops is presented in Fig.\ref{le}. Mean value (Fig. \ref{td11} (d)) of similar pixels within the window can be retrieved from the averaging circuit after the clock signal (Fig. \ref{td11} (c)) activates final flip-flop of the shift register. The reading time, when the mean filter output is activated corresponds to the time period from about 168 ms till 180 ms in Fig.\ref{td11} and Fig.\ref{le}.


\begin{table}[!h]
\centering
\caption{Power dissipation}
\label{tab3}
\begin{tabular}{|l|l|}
\hline
\multicolumn{2}{|c|}{\textbf{Pre-neuron part}}
\tabularnewline
\hline
Difference amplifier                			& 508.9 mW    \\ \hline
VCO                			& $\approx$ 19.1 uW    \\ 
\hline
\multicolumn{2}{|c|}{\textbf{Neuron part}}
\tabularnewline
\hline
Neuron                  		  			& 175  uW         \\ \hline
Comparator                           			& 20 uW           \\ \hline
ADC                         			& 1.6 fW        \\ \hline
Thresholding circuit                     		 	&38 pW      \\ 
\hline
\multicolumn{2}{|c|}{\textbf{Mask storage and filtering}}
\tabularnewline
\hline

Shift register		& 200 pW            \\ \hline

\end{tabular}
\end{table}
\subsection{Discussion}

Presented neuron based filter with the mask size of 3x3 can be incorporated  into CMOS image sensor architecture. Since the proposed filter design requires windowing operation, it is necessary to add two more wire lines to the pixel matrix which brings overall complexity to the circuit design consuming area and power. Nevertheless, embedding the analog filters would be more effective compared to the designs with a separate co-processing unit.  During the system level simulation of proposed filter, its comparison with the Mean filter performance was presented. But comparing it to the existing hardware mean filter designs, the power consumption of existing architecture presented in \cite{soell2015cmos} is less, which is about of 55.3uA current consumption. While proposed method is based on more complex algorithm inspired from the biological neuron functionality, incorporating several computing blocks to the design which causes more power dissipation. Table \ref{tab3} presents the power calculations for each hardware block of the filter. The simulation results show that the rate of power consumption is high only at the stage of defining the similarity of the pixels that relies on differential amplifier requiring 508mW, while the rest of the circuit consume about 220uW  overall. One of the ways to address this problem would be to use low power amplifiers.

\section{Conclusion}
In summary, we presented the design of neuron based adaptive window for image denoising purposes and demonstrated its functionality through simulations on hardware and system levels. The proposed method was verified using Caltech database images and compared to conventional method of noise removing mean filter. It was shown that the neuron can be used for comparison of pixels and serve to build a mask for denoising as well as preserving the information throughout the signal. Hence, incorporation of neuron's behaviour to the architecture of image processing units can further result in self learning systems that can provide better performing results in separating signals from noise. 
 
\balance
      
\bibliographystyle{IEEEtran}
\bibliography{scholar1.bib}

\begin{thebibliography}{10}
\providecommand{\url}[1]{#1}
\csname url@samestyle\endcsname
\providecommand{\newblock}{\relax}
\providecommand{\bibinfo}[2]{#2}
\providecommand{\BIBentrySTDinterwordspacing}{\spaceskip=0pt\relax}
\providecommand{\BIBentryALTinterwordstretchfactor}{4}
\providecommand{\BIBentryALTinterwordspacing}{\spaceskip=\fontdimen2\font plus
\BIBentryALTinterwordstretchfactor\fontdimen3\font minus
  \fontdimen4\font\relax}
\providecommand{\BIBforeignlanguage}[2]{{%
\expandafter\ifx\csname l@#1\endcsname\relax
\typeout{** WARNING: IEEEtran.bst: No hyphenation pattern has been}%
\typeout{** loaded for the language `#1'. Using the pattern for}%
\typeout{** the default language instead.}%
\else
\language=\csname l@#1\endcsname
\fi
#2}}
\providecommand{\BIBdecl}{\relax}
\BIBdecl

\bibitem{v1}
W.-C.~A. Lee, V.~Bonin, M.~Reed, B.~J. Graham, G.~Hood, K.~Glattfelder, and
  R.~C. Reid, ``Anatomy and function of an excitatory network in the visual
  cortex,'' \emph{Nature}, vol. 532, no. 7599, pp. 370--374, 2016.

\bibitem{v2}
R.~M. Cichy, A.~Khosla, D.~Pantazis, A.~Torralba, and A.~Oliva, ``Comparison of
  deep neural networks to spatio-temporal cortical dynamics of human visual
  object recognition reveals hierarchical correspondence,'' \emph{Scientific
  reports}, vol.~6, 2016.

\bibitem{rojas2013neural}
R.~Rojas, \emph{Neural networks: a systematic introduction}.\hskip 1em plus
  0.5em minus 0.4em\relax Springer Science \& Business Media, 2013.

\bibitem{v3}
N.~Spruston, ``Pyramidal neurons: dendritic structure and synaptic
  integration,'' \emph{Nature Reviews Neuroscience}, vol.~9, no.~3, pp.
  206--221, 2008.

\bibitem{gn}
Y.~Akhmetov, J.~J. Mathew, and A.~P. James, ``Variable pixel g-neighbor
  filters.''

\bibitem{chatterjee2010denoising}
P.~Chatterjee and P.~Milanfar, ``Is denoising dead?'' \emph{IEEE Transactions
  on Image Processing}, vol.~19, no.~4, pp. 895--911, 2010.

\bibitem{fpga}
M.~A.~V. Ms. Chipy~Ashok, ``Fpga implementation of image denoising using
  adaptive wavelet thresholding,'' \emph{International Journal of Advanced
  Research in Computer and Communication Engineering}, vol.~4, no.~7, pp.
  288--292, 2015.

\bibitem{sekanina2002image}
L.~Sekanina, ``Image filter design with evolvable hardware,'' in
  \emph{Workshops on Applications of Evolutionary Computation}.\hskip 1em plus
  0.5em minus 0.4em\relax Springer, 2002, pp. 255--266.

\bibitem{vinh2008fpga}
T.~Q. Vinh, J.~H. Park, Y.-C. Kim, and S.~H. Hong, ``Fpga implementation of
  real-time edge-preserving filter for video noise reduction,'' in
  \emph{Computer and Electrical Engineering, 2008. ICCEE 2008. International
  Conference on}.\hskip 1em plus 0.5em minus 0.4em\relax IEEE, 2008, pp.
  611--614.

\bibitem{sun2015cmos}
\BIBentryALTinterwordspacing
J.~Sun, ``Cmos and memristor technologies for neuromorphic computing
  applications, .'' Master's thesis, Technical Report No. UCB/EECS-2015-219,
  EECS Department, University of California, Berkeley, Dec 2015. [Online].
  Available:
  \url{http://www2.eecs.berkeley.edu/Pubs/TechRpts/2015/EECS-2015-219.html}
\BIBentrySTDinterwordspacing

\bibitem{caltech101_2017}
\BIBentryALTinterwordspacing
``Caltech101,'' 2017. [Online]. Available: \url{Vision.caltech.edu}
\BIBentrySTDinterwordspacing

\bibitem{soell2015cmos}
C.~Soell, L.~Shi, A.~Baenisch, T.~Ussmueller, and R.~Weigel, ``A cmos image
  sensor with analog pre-processing capability suitable for smart camera
  applications,'' pp. 279--284, 2015.

\end{thebibliography}

\end{document}